%% file: lineros-taup-proceeding.tex
\documentclass[a4paper]{jpconf}
\usepackage{graphicx}
\usepackage[square,sort&compress]{natbib}
\begin{document}
\title{Galactic positrons and electrons from astrophysical sources and dark matter}

\author{Roberto Lineros}

\address{Dipartimento di Fisica Teorica, Universit\`{a} di Torino and INFN -- Sezione Torino, via P. Giuria 1, I--10125 Torino, Italy.}

\ead{lineros@to.infn.it}

\input{tex/abstr.tex}

\input{tex/intro.tex}

\input{tex/propa.tex}

\input{tex/sourc.tex}

\input{tex/concl.tex}
%

\ack
\begin{small}
Work supported by research grants funded jointly by Ministero dell'Istruzione, dell'Universit\`a e della Ricerca (MIUR), by Universit\`a di Torino (UniTO), by Istituto Nazionale di Fisica Nucleare (INFN) within the Astroparticle Physics Project and by the Italian Space Agency (ASI) under contract N I/088/06/0.
R.L. acknowledges to T. Delahaye, F. Donato, N. Fornengo, J. Lavalle, P. Salati, and R. Taillet for unvaluable comments and suggestions.
\end{small}

\bibliographystyle{iopart-num}
\bibliography{refs-taup}

\end{document}

%% file: tex/abstr.tex
\begin{abstract}
A very interesting puzzle about the origin of electron and positron cosmic rays is deduced from the latests experimental results.
We model the propagation of such cosmic rays in terms of a successfully tested two--zone propagation model.
Several theoretical uncertainties -- like ones related to propagation -- are considered to study different types of electron and positron sources: dark matter annihilation, secondary production, and supernova remnants.\\
\end{abstract}

%% file: tex/intro.tex
\section{Introduction}
\label{sec1}

During last decades, many efforts have been done to understand the nature of cosmic rays. 
Recently, the experimental results for the positron and electron signals obtained by PAMELA~\cite{2009Natur.458..607A}, FERMI~\cite{2009PhRvL.102r1101A}, and HESS~\cite{2008PhRvL.101z1104A} have presented and confirmed a very interesting puzzle related to the origin of such cosmic rays.\\
The positron fraction, the ratio among number of positrons and number of electrons plus positrons, presents an increment for energies above $\sim$50~GeV.
Similar phenomenon appears in the total flux of electron and positron, which presents a change of behavior. For energies below 100~GeV, the total flux is similar to a power--law with a power index of $\sim 3.3$. Above 100~GeV this limit, it changes to a power index of $\sim 3$.\\
In this proceeding, we discuss and study the electron and positron propagation model, different cosmic rays sources as dark matter annihilation and astrophysical ones. 
We focus to study the supernova remnants as a feasible source of electrons.\\

%% file: tex/propa.tex
\section{Electron and positron propagation}
\label{sec2}

The propagation is modeled using a two--zone propagation model~\cite{1980Ap&SS..68..295G}, in which, cosmic rays number density per unit of energy $(\psi)$ is governed by the following transport equation:
\begin{equation}
\label{eq:te}
\frac{\partial \psi }{\partial t} + \nabla \cdot \left(-K_0 \mathcal{\epsilon}^{\delta} \nabla \psi  + \mathbf{V}_c \psi \right) + \frac{\partial J_\epsilon}{\partial \epsilon} = q_{\textnormal{src}} \, .
\end{equation}

For pedagogical purpose, the transport equation~(eq. \ref{eq:te}) is divided into four main parts. 
The first term represents the temporal evolution which is essential to describe propagation when transient sources are present.
The second one describes processes as diffusion due to random magnetic fields -- parameterized by $K_0$ and $\delta$ -- and drift by the galactic wind $\mathbf{V}_c$.
The third one rules the cosmic rays energy losses and gains ($J_{\epsilon}$).
Electrons and positron at energies larger than 10~GeV are dominated mainly by energy losses related to inverse Compton scattering with the interstellar radiation field and synchrotron radiation with galactic magnetic fields. 
The fourth term is the source term which is directly related to electron and positron production processes.\\
The propagation takes place inside a cylinder -- centered in the Galactic Center and oriented like the Galactic Plane -- with a radius set in 20~kpc and half--thickness that typically varies in the range of 1--20~kpc. \\

\subsection{Propagation uncertainties}

Different propagation models, described by different propagation parameter sets, are compatible with cosmic rays observations~\cite{2001ApJ...555..585M};
All these models size the propagation uncertainties.
The analysis of the ratio secondary cosmic rays/primary cosmic rays -- like boron/carbon -- are sensitive to different propagation parameter sets.
Actually, some nuclei cosmic rays are produced mainly by the spallation of primary cosmic rays on the interstellar gas and not by acceleration in supernova remnants.\\
For the study of electron and positron cosmic rays, we use all the boron/carbon compatible parameter sets~\cite{2001ApJ...555..585M}.
The compatible sets serve to size the propagation uncertainties in the electron and positron channel~\cite{2008PhRvD..77f3527D, 2009A&A...501..821D}.\\

%% file: tex/sourc.tex
\section{Sources and fluxes}
\label{sec3}

\begin{figure}[tb]
\centering
\resizebox{0.9\columnwidth}{!}{\includegraphics[angle=270, width=\textwidth]{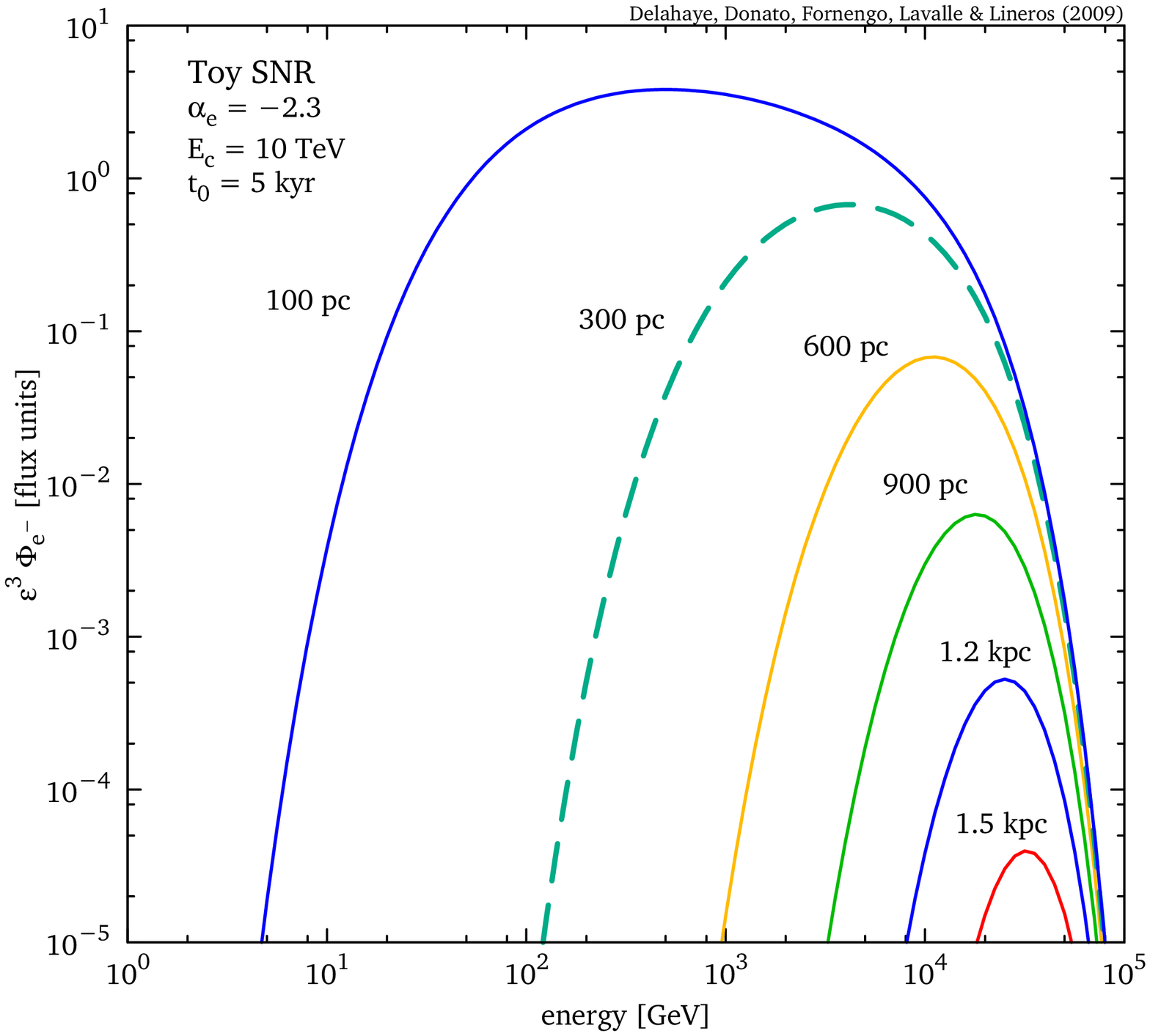} \hspace{3cm}\includegraphics[angle=270, width=\textwidth]{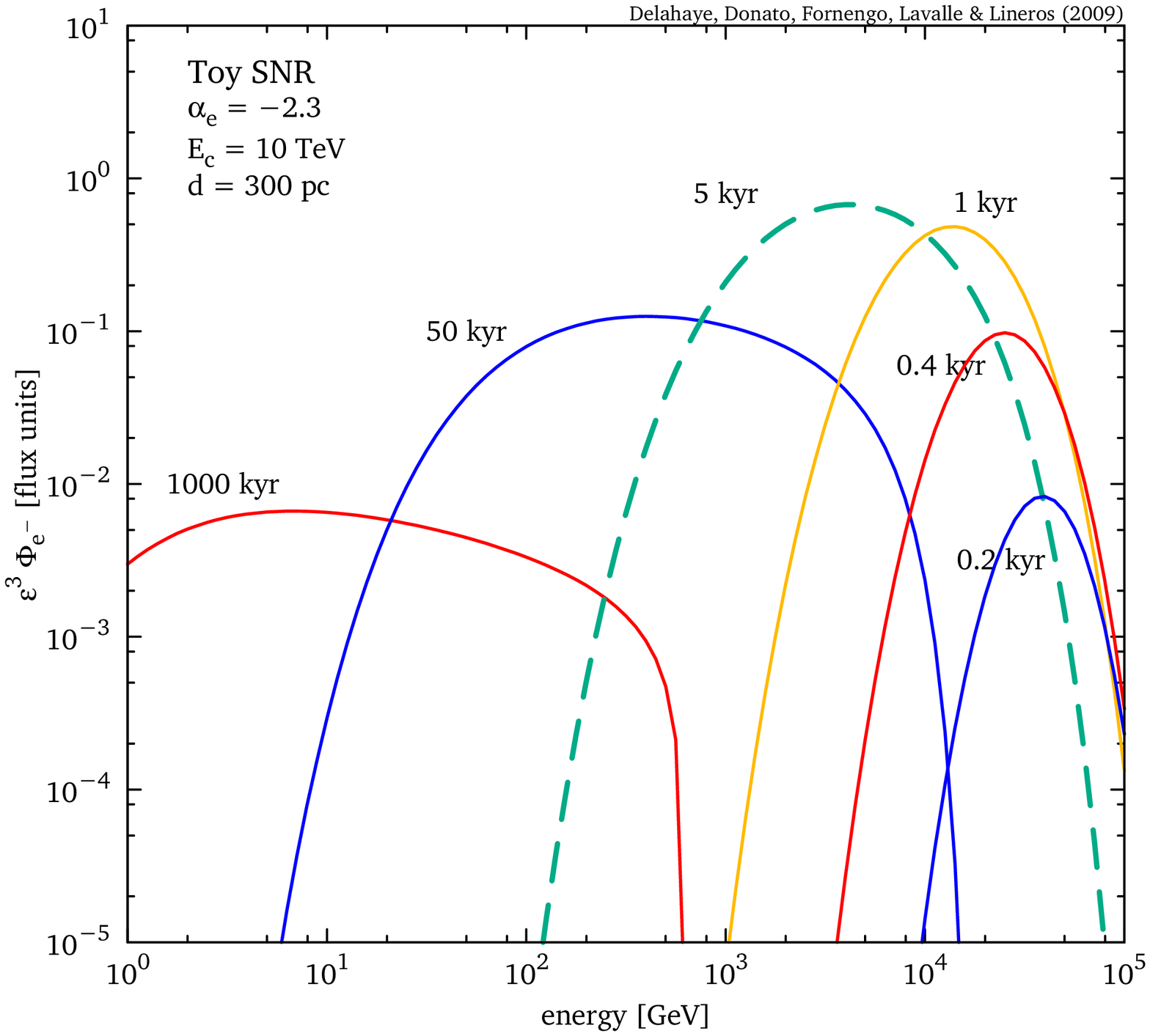}}
\caption{\label{fig1} 
Electron flux $(\varepsilon^3\, \Phi)$ versus energy.
Left panel shows fluxes produced by supernova remnants with same age but at different distances.
Right panel similarly shows cases of remnants with different birth ages and fixed distance to the solar system.
Taking as reference the flux from the remnant at 5~kyr and 300~pc, we observe closer sources contributes more than farther ones. 
On the other hand, older ones contributes at lower energies because particles have lost more energy.
}
\end{figure}

Electron and positron sources can be classified by their origin.
A first studied case is the production from galactic dark matter annihilation~\cite{2008arXiv0812.4272L,2008PhRvD..77f3527D}. 
Electrons, positrons, and other cosmic rays species are not exclusively produced by the dark matter annihilation. 
Generally, those are the result of decays and/or hadronization processes of unstable particles -- like quarks and gauge bosons --  likewise created in the annihilation.
As well, the cosmic rays production depends of the dark matter distribution; Regions where dark matter is denser are regions where annihilation events are more probable.
Another point to consider is that dark matter annihilation corresponds to a type of source which is not localized in the galactic plane, as difference to most of astrophysical sources.
Each discussed point reflects its influence on the propagated electron and positron fluxes when different propagation models are considered~\cite{2008PhRvD..77f3527D}.\\
%
%
Another mechanism to produce electrons and positrons is the interaction between nuclei cosmic rays and the interstellar gas.
We study the secondary production of positron from spallations of proton and alpha particle cosmic rays on hydrogen and helium present in the interstellar medium~\cite{2008arXiv0812.4272L,2009A&A...501..821D}.
The propagated secondary positron flux is more dependent on the propagation uncertainties than other factor, for instance, nuclear cross sections.
Even though, our estimations are encompassed with current observations.
Showing there is consistency among propagation models for nuclei cosmic rays and for electrons and positrons.\\
%
A third type of sources are supernova remnants and pulsars.
Both types are transients sources, in which, most of electrons and positrons are accelerated and injected into the interstellar space in a very short time compared to cosmic rays propagation scales.
A difference with previously discussed sources is that those are rather inhomogeneous in the nearby region to the solar system. 
Usually these kind of source are taken as a smooth distribution across the whole galaxy due to the diffusive propagation of cosmic rays, but when sources are closer enough -- on time and distance -- this approximation is no longer valid~\cite{2009prep}.
In figure \ref{fig1}, we present different examples of supernova remnants at different distances and ages. 
We show how closer remnants contribute more intense to the electron flux. 
Nevertheless, younger ones dominates the high energy part of the flux as consequence of electrons have been affected less time by energy losses.\\

%% file: tex/concl.tex
\section{Conclusions}
\label{sec4}
The improvement in current electron and positron measurements and data has revealed a very interesting puzzle.
Many solution have been proposed during the years.
We present the importance of theoretical uncertainties related to propagation and production of electron and positron cosmic rays~\cite{2008arXiv0812.4272L, 2009A&A...501..821D,2008PhRvD..77f3527D}. 
We stress the necessity to re-estimate secondary and primary component, and to consider already known sources in order to discard/confirm possible presence of an undiscovered component, like the case of dark matter annihilations.
%